\documentclass[12pt, a4paper]{article}
\usepackage[dvipdfmx]{graphicx}
\usepackage{amssymb}
\usepackage{amsmath}
\usepackage{bm}
\usepackage{color}
\usepackage{cite}
\usepackage{slashed}
\usepackage{subfigure}
\usepackage{epstopdf}            
\usepackage{epsfig}
\usepackage{comment}

\newcommand{\beq}{\begin{equation}}
\newcommand{\eeq}{\end{equation}}
\newcommand{\ov}{\overline}

\newcommand{\si}[1]{\textcolor{blue}{#1}}

\usepackage[colorlinks=true, linkcolor=black, citecolor=black,
urlcolor=black]{hyperref}

\begin{document}

\begin{titlepage}

\begin{center}

{\Large{The direct CP violation in a general two Higgs doublet model }}

\vskip 2cm
Syuhei Iguro$^1$ and Yuji Omura$^{2}$  

\vskip 0.5cm

{\it $^1$
Department of Physics, Nagoya University,
Furo-cho Chikusa-ku, Nagoya, Aichi, 464-8602 Japan}\\[3pt]
{\it $^2$ 
Department of Physics, Kindai University, Higashi-Osaka, Osaka 577-8502, Japan}\\[3pt]

\vskip 1.5cm

\begin{abstract}
In this paper, we study the CP violating processes in a general two-Higgs-doublet model (2HDM) with tree-level flavor changing neutral currents. In this model, sizable Yukawa couplings involving top and charm quarks are still allowed by the collider and flavor experiments, while the other couplings are strongly constrained experimentally. The sizable couplings, in general, have imaginary parts and 
could largely contribute to the CP violating observables concerned with the $B$ and $K$ mesons.
In particular, the contribution may be so large that it affects the direct CP violating $K$ meson decay, where the discrepancy between the experimental result and 
the Standard Model prediction is reported. We discuss how well the anomaly is resolved in the 2HDM, based on study of the other flavor observables. We also propose the way to test our 2HDM at the LHC. 
\end{abstract}

\end{center}
\end{titlepage}

\section{Introduction}

The Higgs field is the only fundamental scalar field in the Standard Model (SM).
It has a negative mass squared and spontaneously breaks the electroweak (EW) gauge symmetry, according to the non-vanishing vacuum expectation value (VEV).
The VEV realizes the masses of the SM fermions and the gauge bosons that mediate the 
weak interaction. This scenario can explain almost all of the experimental results,
so that we finally succeed in constructing the model that describes our nature below the EW scale.
The next task is to find out why the SM is so successful in the experiments.
For instance, we do not know the origin of the negative mass squared. 
The Higgs mass squared is unstable against the radiative correction 
and the quartic coupling of the Higgs fields might be driven to the negative value at the high energy by
the radiative correction. 
The origin of the Yukawa couplings between the Higgs field and 
the SM fermions should be also elucidated. It is known that the CP phase in the SM
is not enough to explain the baryon asymmetry in our universe,
so that there would be new physics related to the Higgs mechanism, the Yukawa structure and
the CP violation in our universe. 

There are many possibilities of the new physics to solve the issues concerned with the Higgs field.
Introducing extra symmetry, e.g. supersymmetry and global symmetry,
could make our vacuum stable against the radiative correction.
Extending or unifying the SM gauge symmetry may provide a
natural answer to the question why the mass hierarchy of the down-type quarks
is similar to the one of the charged leptons. Such a unified theory also reveals the 
origin of the SM gauge symmetry \cite{GUT} and could lead the suppressed strong CP phase as well \cite{Babu:1989rb}. We note that extra Higgs fields, that contribute to the EW symmetry breaking,
are predicted at low energy in those scenarios, so that it would be interesting and important
to study the phenomenology relevant to the extra Higgs doublets.
In fact, there are many works on the extended SM with some extra Higgs fields,
motivated by the unified theories \footnote{See, for instance, Refs. \cite{Huitu:1994zm,Frank:2011jia, Frank:2014kma,Babu:2014vba,Dev:2016dja,Iguro:2018oou}.}.
Interestingly, the Higgs fields flavor-dependently couple to
quarks and leptons in some effective models, so that
we can test the unified theories indirectly using the physical observables in flavor-violating processes, even if the scale, where the extended gauge symmetry is broken, is very high \cite{Iguro:2018oou}. 

From the phenomenological point of view, 
it is also very attractive to introduce extra Higgs fields to the SM.
In such a extended SM, we find a rich phenomenology: Higgs physics at the collider experiment,
flavor physics, dark matter physics and so on. In particular, it has been suggested that
the simple extension may solve the anomalies in flavor physics, reported by the
LHCb, Belle and BaBar collaborations \cite{Lees2012xj,Lees2013uzd,Huschle2015rga,Sato2016svk,Hirose2016wfn,Aaij2015yra,Aaij:2013qta,Aaij:2015oid,LHCbnew,Aaij:2014ora,Aaij:2019wad,Abdesselam:2019wac,Abdesselam:2019dgh}.
The authors in Refs. \cite{Ko:2017lzd,Ko:2012sv,Iguro:2017ysu,Crivellin2012ye,Celis:2012dk,Tanaka2012nw,Crivellin:2013wna,Crivellin:2015hha,Cline:2015lqp,
Bian:2017rpg,Bian:2017xzg,Martinez:2018ynq,Iguro:2018fni} have studied the lepton universality of
the semileptonic $B$ decay, $B \to D^{(*)} \ell \nu$ ($\ell=e, \, \mu, \, \tau$), where the excesses are reported in the Belle and the BaBar experiments especially, in a general two-Higgs-doublet model (2HDM). 
In Refs. \cite{Hu:2016gpe,Arnan:2017lxi,Iguro:2018qzf,Li:2018rax}, the anomaly in  $B \to K^{*} \mu^+ \mu^-$ is discussed and the consistency 
with the other flavor-violating processes is investigated. In such a general 2HDM,
there are large tree-level flavor changing neutral currents (FCNCs) involving the two Higgs doublets, so that it is not easy to avoid the constraints from flavor physics.
This general 2HDM can, however, relax the discrepancy between the experimental results
and the SM predictions, in not only the observables mentioned above but also
the anomalous magnetic moment of muon \cite{Omura:2015nja,Omura:2015xcg,Iguro:2018qzf,IOT2}, controlling the tree-level FCNCs by hand\footnote{We do not assume the specific ansatz for the Yukawa structure \cite{Cheng:1987rs,Babu:2018uik}.}. 
There may be new frameworks behind such a phenomenological 2HDM, but, in any case,
it will be very important to investigate the consistency with the latest experimental results
and to survey the way to test the explanation.  
In fact, the direct search for the extra charged scalar at the LHC has begun to exclude the explanation of the anomaly concerned with the lepton universality of the semileptonic $B$ decay \cite{Iguro:2018fni}.
There are many possible underlying theories of a general 2HDM, but it is certain that
the 2HDM with FCNCs can be tested by not only flavor physics but also the direct search for new physics at the LHC even if additional scalars especially couple to fermions in the 3rd generation.

Based on the background and our idea, we investigate the possibility that
the 2HDM can be tested in the CP violation of $K \to \pi \pi$.
The observables concerned with the CP violation in the $K$ decay
have been established. The one measurement is referred to as the indirect CP violation 
and it is induced by the $K-\overline{K}$ mixing. The size of the mixing is given by $\epsilon.$
The other is referred to as the direct CP violation, and the size is given by $\epsilon^\prime$.
The CP violations have been observed and the parameters are measured in the experiments
with high accuracy: $(\epsilon'/\epsilon)_{{\rm exp}}=(16.6\pm2.3)\times10^{-4}$ \cite{Batley:2002gn,AlaviHarati:2002ye,Abouzaid:2010ny} .  On the other hand, it is not easy to obtain the SM predictions because of the strong interaction among the quarks in the hadrons. 
Recently, the lattice QCD calculation of the matrix elements that contribute to the hadronic $K$ decay has been done by the RBC-UKQCD collaboration.
Then, we can obtain the prediction for the direct CP violation.
The SM prediction \cite{Buras:2015yba,Kitahara:2016nld} with RBC-UKQCD lattice collaboration \cite{Bai:2015nea,Blum:2015ywa} and the Dual QCD approach (DQCD)\cite{Buras:2015xba,Buras:2016fys} is given as
\beq
\label{DQCD}
(\epsilon'/\epsilon)_{SM}=((1-2)\pm5)\times10^{-4}.
\eeq
There is a discrepancy between this result and the experimental result.
The authors of \cite{Gisbert:2017vvj}, on the other hand, obtain $(\epsilon'/\epsilon)_{SM}=(15\pm7)\times10^{-4}$ with ideas from the chiral perturbation theory, that agrees with the experimental value.
It would be too early to conclude that the discrepancy is from new physics, but
it would be important to summarize the new physics possibilities taking into 
consideration the consistency with the other observables in the new physics.
We can actually find some new physics candidates \cite{Buras:2015kwd,Tanimoto:2016yfy,Buras:2016dxz,Endo:2016aws,Bobeth:2016llm,Endo:2016tnu,Bobeth:2017xry,Crivellin:2017gks,Haba:2018byj,Haba:2018rzf}.
In this paper, we investigate how well a general 2HDM can fit the experimental result
of the direct CP violation. 
In a general 2HDM, there are many experimental constraints
since sizable flavor-violating Yukawa couplings between the SM fermions and the extra scalars
are allowed. The couplings that can be sizable from the phenomenological point of view
are the ones involving charm and top quarks \cite{Altunkaynak:2015twa}.
Then, we concentrate on the contribution of the sizable couplings to
the direct CP violation, and discuss the consistency with not only the other flavor observables
but also the measurements at the LHC.
Compared to the previous work \cite{Chen:2018ytc}, we also take into account the constraint from the CP violation
in $B$ meson mixing and survey the heavy mass region and CP phases in the Yukawa coupling as well. We also discuss signals at the LHC to test our model.

In Sec. \ref{sec2}, we introduce our general 2HDM.
In Sec. \ref{sec3}, we discuss our prediction for the direct CP violation and
summarize the relevant experimental constraints.
Based on the result in Sec. \ref{result}, we propose the way to
test our model at the LHC in Sec. \ref{sec4}.
Sec. \ref{summary and comment} is denoted to the summary.


\section{General two Higgs doublet model}
\label{sec2}

First of all, we introduce our 2HDM with tree-level FCNCs.
There are two Higgs fields charged under the EW symmetry, $SU(2)_L \times U(1)_Y$,
and both fields generally contribute to the EW symmetry breaking.
To see the mass eigenstates of the scalars clearly, 
let us define the two Higgs fields in the base where only one of the Higgs doublets
develops the VEV.  This base is known as a Higgs basis or Georgi basis \cite{Georgi:1978ri,Donoghue:1978cj}, and those two Higgs doublets in this base are written as 
\begin{eqnarray}
  H_1 =\left(
  \begin{array}{c}
    G^+\\
    \frac{v+\phi_1+iG}{\sqrt{2}}
  \end{array}
  \right),~~~
  H_2=\left(
  \begin{array}{c}
    H^+\\
    \frac{\phi_2+iA}{\sqrt{2}}
  \end{array}
  \right),
\label{HiggsBasis}
\end{eqnarray}
where $G^+$ and $G$ are Nambu-Goldstone bosons, and $H^+$ and $A$ are physical mass eigenstates called a charged Higgs boson and a CP-odd
Higgs boson, respectively. $v$ is the non-vanishing VEV and satisfies $v \simeq 246$ GeV.
The CP-even neutral Higgs bosons, $\phi_1$ and $\phi_2$, mix each other and lead mass
eigenstates, $h$ and $H$ as follows:
\begin{eqnarray}
  \left(
  \begin{array}{c}
    \phi_1\\
    \phi_2
  \end{array}
  \right)=\left(
  \begin{array}{cc}
    \cos\theta_{\beta \alpha} & \sin\theta_{\beta \alpha}\\
    -\sin\theta_{\beta \alpha} & \cos\theta_{\beta \alpha}
  \end{array}
  \right)\left(
  \begin{array}{c}
    H\\
    h
  \end{array}
  \right).
\end{eqnarray}
Here $\theta_{\beta \alpha}$ is the mixing angle and we identify $h$ as the scalar with 125 GeV mass and assume that $H$ is heavier than $h$. Then, the coupling of $h$
becomes identical to the SM one in the limit, $\sin\theta_{\beta\alpha}\rightarrow 1$.

In our model, we do not assign any symmetry to distinguish the two Higgs fields,
so both Higgs doublets can couple to all fermions.
Then, the Yukawa interactions between the mass eigenstates of the SM fermions and the Higgs fields are given by
\begin{eqnarray}
  {\cal L} &=&-\bar{Q}_L^i H_1 y_d^i d_R^i -\bar{Q}_L^i H_2 \rho_d^{ij} d_R^j-\bar{Q}_L^i (V^\dagger)^{ij} \widetilde{H_1} y_u^j u_R^j
  -\bar{Q}_L^i (V^\dagger)^{ij} \widetilde{H_2} \rho_u^{jk}u_R^k \nonumber\\
  &&-\bar{L}_L^i H_1 y^i_e e_R^i -\bar{L}_L^i H_2 \rho^{ij}_e e_R^j.
\label{yukawas}
\end{eqnarray}
Here the indices, $i$, $j$ and $k$, represent flavor. $\widetilde{H_{1,2}}=i \tau_2 H^*_2$ is defined using the Pauli matrix, $\tau_2$.
The $SU(2)_L$ doublet quarks and leptons are defined as $Q=(V^\dagger u_L,d_L)^T$ and $L=(V_{\rm MNS} \nu_L, e_L)^T$, using the CKM matrix, $V$, and the MNS matrix, $V_{\rm MNS}$.
Note that Yukawa couplings $y^i_f$ are expressed by the fermion masses $m_{f^i}$ as
$y^i_f=\sqrt{2}m_{f^i}/v$ $(f=d, \, u, \, e)$, that are the same as the SM Yukawa couplings.
On the other hand, Yukawa couplings $\rho_f$ are unknown general $3\times3$ complex matrices and sources of the Higgs-mediated flavor violation.

In mass eigenstates of Higgs bosons, the Yukawa interactions are given by
\begin{align}
  {\cal L}&=-\sum_{f=u,d,e}\sum_{\phi=h,H,A} y^f_{\phi i j}\bar{f}_{Li} \phi f_{Rj}+{\rm h.c.}
  \nonumber\\
  &\quad -\bar{\nu}_{Li} (V_{\rm MNS}^\dagger \rho_e)^{ij}  H^+ e_{Rj}
  -\bar{u}_i(V\rho_d P_R-\rho_u^\dagger V P_L)^{ij} H^+d_j+{\rm h.c.},
\end{align}
where
\begin{align}
  y^f_{hij}=\frac{m_{f}^i}{v}s_{\beta\alpha}\delta_{ij}&+\frac{\rho_{f}^{ij}}{\sqrt{2}} c_{\beta\alpha},~~
  y^f_{Hij}=\frac{m_f^i}{v} c_{\beta \alpha}\delta_{ij}-\frac{\rho_f^{ij}}{\sqrt{2}} s_{\beta\alpha},
  \nonumber \\
  y^f_{Aij}&=
  \left\{
  \begin{array}{c}
    -\frac{i\rho_f^{ij}}{\sqrt{2}}~({\rm for}~f=u),\\
    \frac{i\rho_f^{ij}}{\sqrt{2}}~({\rm for}~f=d,~e),
  \end{array}
  \right.
  \label{yukawa}
\end{align}
where $c_{\beta\alpha}$ and $s_{\beta\alpha}$ are short for $\cos\theta_{\beta\alpha}$ and $\sin\theta_{\beta\alpha}$ respectively.
Note that the Yukawa interactions of $h$ become SM-like when $c_{\beta\alpha}$ is small, but there are small flavor-violating interactions $\rho_f^{ij}$ suppressed by $c_{\beta\alpha}$. On the other hand, the Yukawa interactions of heavy Higgs bosons ($H$, $A$, and $H^+$)
are mainly controlled by the $\rho_f$ couplings, so that they can be sizable as far as the experimental constraints are not so tight.
We consider the case $c_{\beta\alpha}=0$ for simplicity.
It is known that the elements of $\rho_u$ involving charm and top quarks 
can be sizable since they are not strongly constrained by the flavor physics.
Besides, it is pointed out that the sizable $\rho_u^{tc}$  can improve the discrepancies in the $b \to s ll$ and $b \to c l \nu$ processes \cite{Iguro:2017ysu,Iguro:2018qzf}. Based on these previous works on the 2HDM, we consider the following simple textures of $\rho_f$ from the phenomenological point of view: 
\begin{equation}
\label{eq;rhou}
\rho_u \simeq \begin{pmatrix} 0 & 0 & 0 \\ 0 & \rho_u^{cc} & \rho_u^{ct} \\ 0 &   \rho_u^{tc} & \rho_u^{tt}  \end{pmatrix},~|\rho_d^{ij}| \ll {\cal O} (0.1),~|\rho_e^{ij}| \ll {\cal O} (0.1).
\end{equation}
The other elements of $\rho_u$ are assumed to be less than ${\cal O} (0.01)$,
so that the physics involving $\rho_u^{ct}$, $\rho_u^{tc}$, and $\rho_u^{tt}$ is mainly discussed in this paper.

Before our study of the phenomenology, let us summarize the masses of the heavy scalars.
They are evaluated as
\begin{eqnarray}
  m_H^2& \simeq& m_A^2+\lambda_5 v^2, \\
  m_{H^\pm}^2 &\simeq& m_A^2-\frac{\lambda_4-\lambda_5}{2} v^2.
  \label{Higgs_spectrum}
\end{eqnarray}
$m_H$, $m_A$ and $m_{H^+}$ denote the masses of the heavy CP-even, CP-odd and charged Higgs scalars, and
$\lambda_4$ and $\lambda_5$ are the dimensionless couplings in the Higgs potential:
$V(H_i)=\lambda_4 (H_1^\dagger H_2)(H_2^\dagger H_1) + \frac{\lambda_5}{2}(H_1^\dagger H_2)^2+ \dots $
In order to evade the stringent bound from the EW precision observables (EWPOs) \cite{Peskin:1990zt}, we assume that the scalar masses are degenerate: $m_H=m_A=m_{H^+}$.\footnote{We note that
our result of the flavor physics does not change even if the mass difference is sizable, since the charged scalar only contributes to the flavor physics in our analysis.}

\section{Phenomenology}
\label{sec3}
In this section, we discuss phenomenology: flavor physics and the signals at the LHC.
First, we study the direct CP violation in the $K$ meson decay.
\subsection{The direct CP violation in $K \to \pi \pi$}
\subsubsection{Model-independent analysis}
The $K$ decay, $K \to \pi \pi$, has been well studied motivated by the CP violation.
There are two types of the CP violation: one is indirect and the other is direct.
The latter contribution is described by the $|\Delta S|=1$ effective Hamiltonian expressed as \cite{Kitahara:2016nld}
\begin{align}
\label{eq;Heff}
{\cal{H}}_{eff}^{|\Delta S|=1}=\frac{G_F}{\sqrt 2} \sum^{10}_{i=1} Q_i(\mu) s_i(\mu)+h.c.,
\end{align}
where the effective operators are defined as \cite{Buras:1993dy}
\begin{align}
&\rm{{\bf{Current-Current}}}\nonumber\\
Q_1&=(\ov{s_\alpha}u_\beta)_{V-A}(\ov{u_\beta} d_\alpha)_{V-A},~~~Q_2=(\ov{s}u)_{V-A}(\ov{u} d)_{V-A},\nonumber\\
&\rm{{\bf{QCD ~Penguin}}}\nonumber\\
Q_3&=(\ov{s}d)_{V-A}\sum_q(\ov{q} q)_{V-A},~~~Q_4=(\ov{s_\alpha}d_\beta)_{V-A}\sum_q(\ov{q_\beta} q_\alpha)_{V-A},\nonumber\\
Q_5&=(\ov{s}d)_{V-A}\sum_q(\ov{q} q)_{V+A},~~~Q_6=(\ov{s_\alpha}d_\beta)_{V-A}\sum_q(\ov{q_\beta} q_\alpha)_{V+A},\nonumber\\
&\rm{{\bf{EW ~Penguin}}}\nonumber\\
Q_7&=\frac{3}{2}(\ov{s}d)_{V-A}\sum_q e_q (\ov{q} q)_{V+A},~~~Q_8=\frac{3}{2}(\ov{s_\alpha}d_\beta)_{V-A}\sum_q e_q(\ov{q_\beta} q_\alpha)_{V+A},\nonumber\\
Q_9&=\frac{3}{2}(\ov{s}d)_{V-A}\sum_q e_q (\ov{q} q)_{V-A},~~~Q_{10}=\frac{3}{2}(\ov{s_\alpha}d_\beta)_{V-A}\sum_q e_q(\ov{q_\beta} q_\alpha)_{V-A}.
\label{Operators}
\end{align}
We note that $V\pm A=\gamma_\mu(1\pm \gamma_5)$, $\alpha$ and $\beta$ are color induces,  and $e_q$ is a electric charge of q quark.

The mass eigenstates of the neutral $K$ mesons are given by the linear combination of $K$ and $\overline{K}.$ One is a long-lived meson named $K_L$ and the other is named $K_S$ that has a short lifetime. $K_L$ ($K_S$) is close to the CP-odd (CP-even) state, so that $K_L$ dominantly decays to three $\pi$ while $K_S$ decays to two $\pi$. The measurement concerned with the CP violation is given by surveying the decay of $K_L \to 2 \, \pi$.  

The two $\pi$ in the final state of $K \to 2 \, \pi$ consist of the isospin-zero ($I=0$) state and the isospin-two ($I=2$) state. Taking the ratio of the amplitude of $K_L \to (\pi \pi)_{I=0}$
to the one of $K_S \to (\pi \pi)_{I=0}$, $\epsilon_K$ that is measurement of the indirect CP violation is obtained.

The measurement of the direct CP violation is given by taking into account the decay to the $I=2$ state: $K_L \to (\pi \pi)_{I=2}$ and $K_S \to (\pi \pi)_{I=2}$. When we define the amplitudes of $K \to (\pi \pi)_{I=0, \, 2}$ as ${\rm A}_{0, 2}$, the parameter, $\epsilon^\prime/\epsilon$, to measure the direct CP violating decay is given by \cite{Buras:2015jaq}
\begin{align}
\frac{\epsilon'}{\epsilon}=-\frac{\omega}{\sqrt {2} |\epsilon_K|}\left[ \frac{\rm{Im}A_0}{\rm{Re}A_0}(1-\Omega_{eff})-\frac{1}{a}\frac{\rm{Im}A_2}{\rm{Re}A_2}\right].
\end{align}
In our study, we approximately evaluate $\epsilon^\prime/\epsilon$ as follows.
The parameter relevant to the  indirect CP violation is fixed at the well measured experimental value: $|\epsilon_K|=(2.228(11))\times10^{-3}$.
$a$ and $\Omega_{eff}$ describe the isospin breaking effects:
 $a=1.017$ and $\Omega_{eff}= (14.8 \pm 8.0) \times 10^{-2}$ \cite{Buras:2015yba,Cirigliano:2003nn}.
$\omega$  is the ratio of ${\rm{Re}A_2}$ to ${\rm{Re}A_0}$ and fixed at $\omega=(4.53\pm0.02)\times10^{-2}$ in our analysis.
 Following the analysis in Ref. \cite{Kitahara:2016nld}, we obtain the evaluation of $\epsilon^\prime/\epsilon$ as
\begin{eqnarray}
\frac{\epsilon'}{\epsilon}&=&\left ( \frac{\epsilon'}{\epsilon} \right )_{\rm{SM}}+\left (  \frac{\epsilon'}{\epsilon} \right )_{\rm{H}},  \\
\left (  \frac{\epsilon'}{\epsilon} \right )_{\rm{H}} & \simeq & \frac{G_F \, \omega}{2|\epsilon_K| \, {\rm{Re}A_0}}\times \left \{\sum_{i,j} \langle Q_i(\mu) \rangle  \, U_{ij}(\mu, \mu_{NP}) \, {\rm{Im}} ~s_j(\mu_{NP}) \right \}.
\end{eqnarray}
$(\epsilon' / \epsilon)_{\rm{SM}}$ and $(\epsilon' / \epsilon)_{\rm{H}}$ denote the SM contribution and the extra Higgs contribution, respectively.
$U_{ij}$ takes into account the renormalization-group (RG) correction to $s_i$, which is a short-distance correction, and $\langle Q_i(\mu) \rangle$ corresponds to the hadronic matrix element.
We use the result of the RBC-UKQCD lattice collaboration \cite{Bai:2015nea,Blum:2015ywa}
for the evaluation of the matrix elements and obtain the numerical values
for $ F_j (\mu_{NP})= \langle Q_i(\mu) \rangle U_{ij}(\mu, \mu_{NP}) $, based on Ref. \cite{Kitahara:2016nld}.
When we fix the new physics scale at $\mu_{NP}=1$TeV, we find the numerical evaluation as
\begin{align}
&F_3(1\rm{TeV})=0.045,~F_4(1\rm{TeV})=-0.193,~F_5(1\rm{TeV})=0.081,~F_6(1\rm{TeV})=0.305,\nonumber\\
&F_7(1\rm{TeV})=26.16,~F_8(1\rm{TeV})=88.61,~F_9(1\rm{TeV})=0.117,~F_{10}(1\rm{TeV})=-0.084.
\end{align}
We use these values to study the new physics effect to the direct CP violation.


\subsubsection{Prediction of the general 2HDM for $\epsilon'/\epsilon$}
\label{epepG2HDM}
We study the contribution of the heavy charged scalar to $\epsilon'/\epsilon$.
We note that the diagram involving neutral heavy scalars is suppressed as far as $\rho_d$
is vanishing. If the off-diagonal elements of $\rho_d$ are sizable, we suffer from the tree-level FCNCs so that
we consider the case that the off-diagonal elements of $\rho_u$ involving heavy quarks are only sizable,
assuming the Yukawa alignment as in Eq. (\ref{eq;rhou}).
\\
\begin{figure}[t]
  \begin{center}
    \includegraphics[width=5cm]{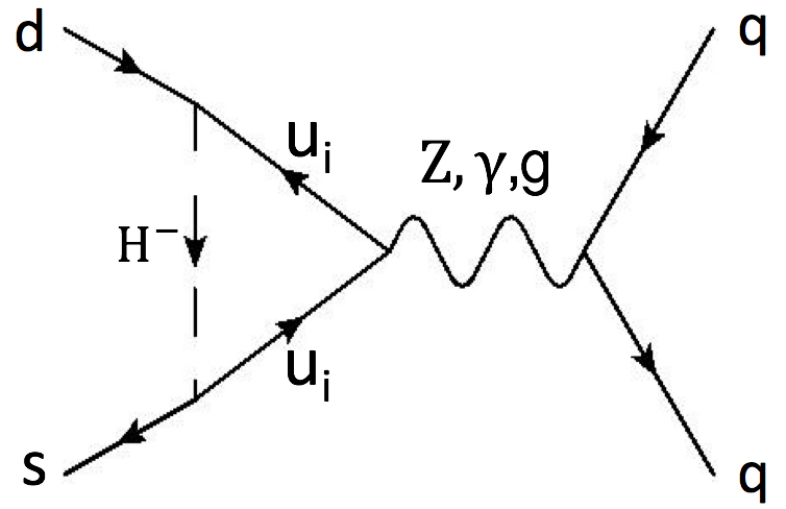}
            \caption{The one-loop penguin diagrams relevant to $\epsilon'/\epsilon$.}
    \label{diaBP}
  \end{center}
\end{figure}
The one-loop penguin diagrams involving charged heavy scalar are shown in Fig. \ref{diaBP}.
These diagrams dominantly contributes to $\epsilon'/\epsilon$, while
the box diagram is subdominant numerically and can be neglected.
\footnote{The box diagram is induced by the charged scalar and W boson exchanging, but it is suppressed by up-type quark masses and the CKM matrix.}
Then the coefficients in Eq. (\ref{eq;Heff}) are
given by the photon-penguin, $Z$-penguin, and gluon-penguin diagrams
that are defined as $s^\gamma_i$, $s^Z_i$ and $s^g_i$, respectively:
\beq
s_i=s^\gamma_i+s^Z_i+s^g_i.
\eeq
The each contribution is described as
\begin{align}
s_7^{\gamma}&=s_{9}^{\gamma}=-\frac{\alpha}{6 \pi}C_{\gamma},\\
s_3^{g}&=s_5^{g}= \frac{\alpha_s}{24 \pi} C_{g},\\
s_4^{g}&=s_6^{g}=-\frac{\alpha_s}{8 \pi} C_{g},\\
s_3^{Z}&=\frac{1}{48\pi^2} C_{Z},  \\
~~s_7^{Z}&=\frac{s^2_w}{12\pi^2} C_{Z} ,\\
~~s_9^{Z}&=-\frac{1-s_w^2}{12\pi^2} C_{Z} ,
\end{align}
where $s_w$ is short for the sine of the Weinberg angle and the functions in the right-hand side are defined with $x_i=m_{u_i}^2/m_{H}^2$ as
\begin{align}
     C_{\gamma} =&\frac{1}{2\sqrt{2} G_{\rm F}m_{H^+}^2 }
     \sum_i (V^\dagger \rho_u)^{si}(\rho_u^\dagger V)^{id}
    \left[ \frac{2}{3} G_{\gamma 1}(x_i)+ G_{\gamma 2}(x_i)\right], \\
    C_{g} =&\frac{1}{2\sqrt{2} G_{\rm F}m_{H^+}^2 }
     \sum_i (V^\dagger \rho_u)^{si}(\rho_u^\dagger V)^{id}
    \left[ G_{\gamma 1}(x_i)\right], \\
    C_{Z} =&\sum_i (V^\dagger \rho_u)^{si}(\rho_u^\dagger V)^{id} G_Z(x_i).
    \label{C9_general}  
\end{align}
Here the functions are defined as
\begin{align}
G_{\gamma 1}(x)&=-\frac{16-45x+36x^2-7x^3+6(2-3x)\log x}{36 (1-x)^4},\\
G_{\gamma 2}(x)&=-\frac{2-9x+18x^2-11x^3+6x^3\log x}{36 (1-x)^4},\\
G_Z(x) &=\frac{x(1-x+\log x)}{2(1-x)^2}.
\end{align}

We note that there are dipole operators that contribute to
the $\Delta S=1$ processes, but they are very much suppressed by our $\rho_u$ alignment.
Therefore, we focus on the penguin-diagram contribution discussed above in our analysis.


\subsection{The other constraints from flavor physics}
In this subsection, we summarize the constraints from other flavor-violating processes in our model. The sizable Yukawa couplings between the heavy quarks, $c$ and $t$, and
the heavy (charged) scalars significantly contribute to
$b\to s\gamma$,  $B_{d(s)}-\overline{B_{d(s)}}$ mixing and $|\epsilon_K|$.
The constraints on our model from those observables have been studied by the authors in Refs. \cite{Iguro:2018qzf, Iguro:2018oou,Chen:2018ytc,Chen:2018vog}. In addition to the results of the previous works,
we take into account the constraint from the CP violation in the $B_{d(s)}-\overline{B_{d(s)}}$ mixing. When we define the relevant parameters as
\begin{equation}
C_{B_q}e^{2i\phi_{B_q}}=\frac{\langle B_q|H_{eff}^{full}|\overline{B_q}\rangle}{\langle B_q|H_{eff}^{SM}|\overline{B_q}\rangle},
\end{equation}
the parameters to fit the experimental results are introduced by the UTfit collaboration \cite{UTfit}:
\begin{align}
C_{B_d}=&1.05\pm0.11,~~\phi_{B_d}=-2.0\pm1.7,\\
C_{B_s}=&1.110\pm0.090.
\end{align}
Using those values, we draw the exclusion lines allowing the 95$\%$ CL interval from the central values. $\phi_{B_d}$ is referred to as a golden mode and can be measured precisely, while $\phi_{B_s}$ is not measured as accurately as $\phi_{B_d}$ \cite{Tanabashi:2018oca}. Then, we do not include $\phi_{B_s}$ in our analysis.

Besides, we consider the constraint from $\epsilon_K$.
We require that the prediction is within the SM prediction with 2$\sigma$ errors of $\eta_{1,2,3}.$
We also checked that the constraints from $b\to s\gamma$ is less stringent.
In addition, we found that the bound from $BR(K_L\to\mu\bar\mu)$ is less stringent \cite{Buras:2015yca,Isidori:2003ts}. 

\subsection{Predictions for $BR(K\to\pi \nu\bar\nu)$ and $BR(B_s\to \mu\bar\mu)$ }
Before showing our numerical results, let us discuss our predictions for $BR(K\to\pi \nu\bar\nu)$ and $BR(B_s\to \mu\bar\mu)$. Those observables are expected to be measured with high accuracy in the near future, so that
they may play a crucial role in testing our model.

In the rare $K$ meson decay, there are two processes: $K^+\to\pi^+ \nu\bar\nu$ and $K_L\to\pi^0 \nu\bar\nu$.
They are investigated in the NA62 and KOTO experiments, respectively \cite{Lurkin:2018gdo},
so that we look into some explicit correlations between $BR(K\to\pi \nu\bar\nu)$ and the direct CP violation in our model.
The $K \to\pi \nu\bar\nu$ decay is given by the following four-fermi operator \cite{Buchalla:1998ba},
\begin{align}
{\cal H}^{\nu}_{eff}=\frac{G_F\alpha}{2\sqrt 2 \pi s_w^2}V^*_{ts}V_{td} \, X \, (\bar {s} d)_{V-A}(\bar \nu \nu)_{V-A}+h.c..
\end{align}
$X$ includes both the SM and the 2HDM contributions.
In our general 2HDM, the following operator is effectively generated by 
the penguin diagram involving the charged scalar: 
\begin{align}
\Delta {\cal H}^{\nu}_{eff}&=-\frac{G_F}{2 \sqrt{2}} \frac{C_Z}{16 \pi^2} (\bar {s} d)_{V-A}(\bar \nu \nu)_{V-A}+h.c.,
\end{align}
where $C_Z$ is given in Eq. (25).
Then, the contribution of the 2HDM to $X$ is described as
\begin{eqnarray}
X &=& X_{SM}+ \Delta X,  \\
\Delta X &=& -\frac{s_w^2 C_Z}{8\pi\alpha V_{td}V_{ts}^*},
\end{eqnarray}
where $X_{SM}$ denotes the SM contribution.
BR($K\to\pi \nu\bar\nu$) is calculated with $X$ as 
\begin{align}
BR(K^+\to\pi^+ \nu\bar\nu)&=k_+(1+\Delta_{EM})\left[ \left( \frac{\rm{Im}[\lambda_t X]}{\lambda^5}\right)^2+\left(\frac{\rm{Re}[\lambda_c]}{\lambda}P_c+\frac{\rm{Re}[\lambda_t X]}{\lambda^5} \right)^2\right],\\
BR(K_L\to\pi^0 \nu\bar\nu)&=k_L \left( \frac{\rm{Im}[\lambda_t X]}{\lambda^5}\right)^2.
\end{align}
To evaluate our prediction, we use $k_+=(5.173\pm0.025)\times10^{-11}\left(\frac{\lambda}{0.225}\right)^8$, $k_L=(2.231\pm0.013)\times10^{-10}\left(\frac{\lambda}{0.225}\right)^8$, $P_c=0.404\pm0.024$,  $\Delta_{EM}=-0.003$, $X_{SM}=1.481\pm 0.009$,
$m_c(m_c)=1.275$ GeV and $m_t(m_t)=160$ GeV\cite{Buras:2015qea}.

The rare leptonic $B_s$ decay, $B_s\to\mu\bar\mu$, is one of the main targets of the LHC experiment.
$BR(B_s\to\mu\bar\mu)$ is measured as  $BR(B_s\to\mu\bar\mu)_{{\rm exp}}=(2.7\pm0.6)\times10^{-9}$ \cite{Tanabashi:2018oca}, and can be expected to be measured with high accuracy in the future.
In our model, the Z-penguin diagram contributes to the decay as, 
\begin{align}
BR(B_s\to\mu\bar\mu)=BR(B_s\to\mu\bar\mu)_{SM}\left |1+\frac{C_{10}^{G2HDM}}{C_{10}^{SM}}\right|^2.
\end{align}
We note that $BR(B_s\to\mu\bar\mu)_{SM}$ is $3.32\times10^{-9}$ in our study.
Following Ref. \cite{Iguro:2017ysu}, we calculate our prediction for this leptonic $B_s$ decay.

\subsection{Numerical result}
\label{result}

Based on the above discussion, we numerically analyze our model.
We mainly focus on the contribution of $\rho_u^{tt}$, $\rho_u^{tc}$, $\rho_u^{ct}$ and $\rho_u^{cc}$
that are the Yukawa couplings involving heavy quarks. In particular, top and charm quarks only appear
in the one-loop diagram relevant to $B$ and $K$ physics:
the diagram induced by $\rho_u^{ct}$ and $\rho_u^{tt}$ involves top quarks in the loop, while
the diagram induced by $\rho_u^{tc}$ and $\rho_u^{cc}$ involves charm quarks.
In order to study the each contribution quantitatively, 
we consider the following two cases:
\begin{enumerate}
\centering
\renewcommand{\labelenumi}{(\roman{enumi})}
\item[(I)] $\left | \rho_u^{tt} \right |$, $\left | \rho_u^{ct} \right |$ $\gg$ $\left | \rho_u^{tc}\right |$, $ \left | \rho_u^{cc}\right |,$
\item[(II)] $\left | \rho_u^{tt} \right |$, $\left | \rho_u^{ct} \right |$ $\ll$ $\left | \rho_u^{tc}\right |$, $ \left | \rho_u^{cc}\right |$.
\end{enumerate}
In the case (I) (case (II)), top quark (charm quark) runs in the penguin diagram in Fig. \ref{diaBP}.

First, we discuss the case (I). In this case, the dominant contribution to the direct CP violation comes from the one-loop diagram with top quark that is linear to $\rho_u^{tt}\times\rho_u^{ct}$. The relevant parameters are $m_H$, $|\rho_u^{ct}|$, $|\rho_u^{tt}|$ and a relative phase between $\rho_u^{ct}$ and $\rho_u^{tt}$.
In our presentation, we set $\rho_u^{tt}$ to unit
and look into the $\rho_u^{ct}$ dependence on the flavor physics.
 \begin{figure}[t]
  \begin{center}
      \includegraphics[width=7.4cm]{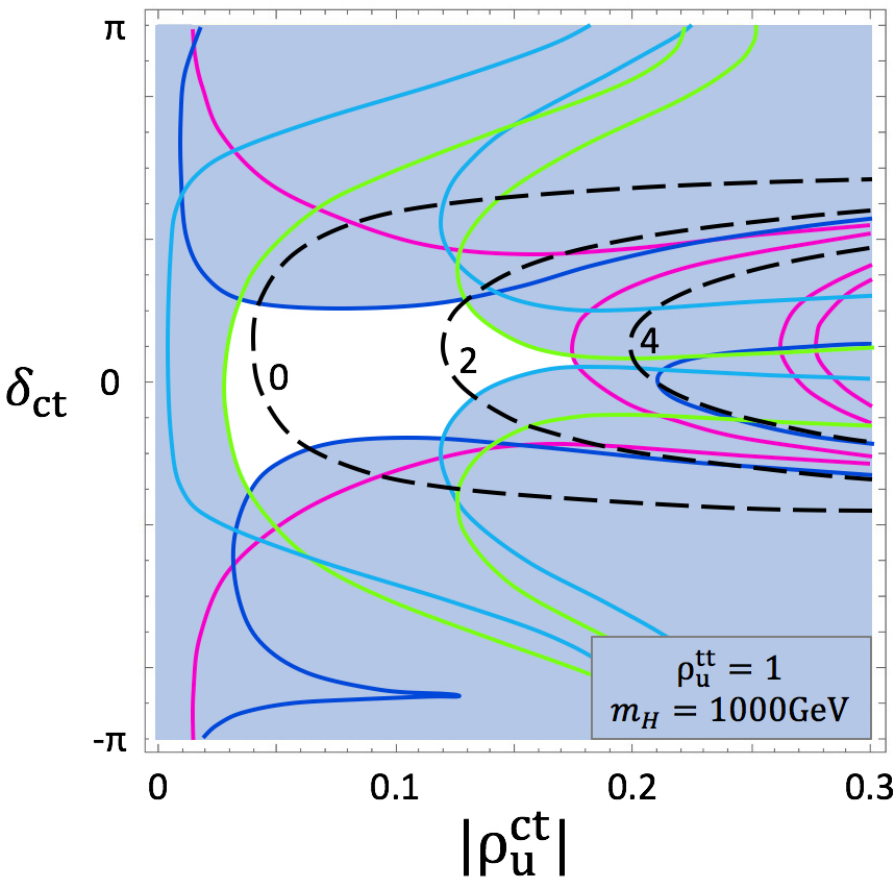}
            \includegraphics[width=7.6cm]{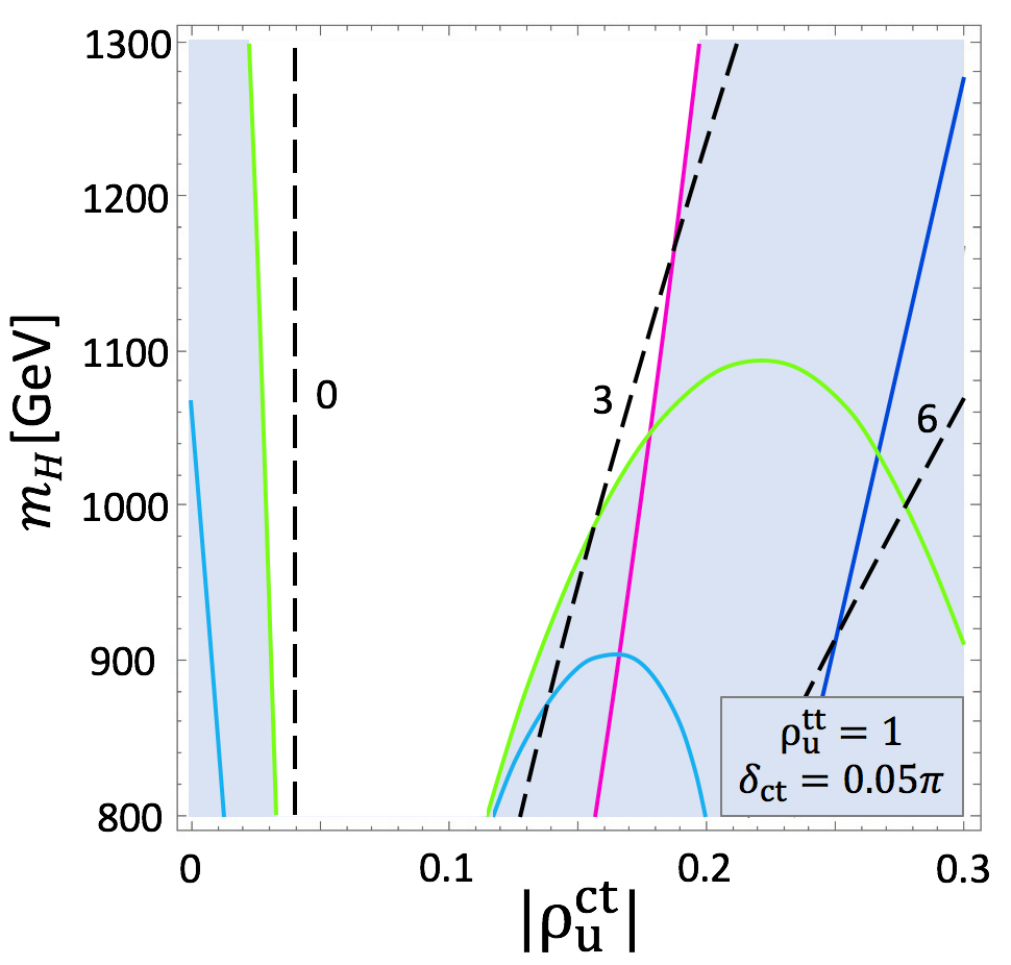}
        \caption{ The prediction for $\epsilon'/\epsilon_{\rm{H}}\times10^4$ is expressed in dashed black lines in the case (I). In each figure, the exclusion lines explained in the text are shown: $|\epsilon_K|$ (magenta), $C_{B_d}$ (cyan), $C_{B_s}$ (light green) and $\phi_{B_d}$ (blue).}
    \label{re2}
  \end{center}
\end{figure}

Fig. \ref{re2} shows the prediction for $(\epsilon'/\epsilon)_{{\rm H}}$ in the case (I). 
The prediction is depicted by the dashed black lines with the number that corresponds to $(\epsilon'/\epsilon)_{{\rm H}} \times10^4$.
In the left panel, the vertical and horizontal axes denote the phase and the absolute value of $\rho_u^{ct}$ respectively. We define the phase as $\rho_u^{ct}=|\rho_u^{ct}| e^{i\delta_{ct}}$
and charged scalar mass, $m_H$, is fixed at $1$ TeV.
In the right panel, the vertical and horizontal axes denote the heavy Higgs mass and the absolute value of $\rho_u^{ct}$ respectively. The phase is fixed at $\delta_{ct}=0.05\, \pi$ that drastically relaxes the bound from the $B$ meson mixing.
In the each panel, the exclusion lines of $|\epsilon_K|$, $C_{B_d}$, $C_{B_s}$ and $\phi_{B_d}$ are drawn by the solid lines colored in magenta, cyan, light green and blue, respectively. 
The shaded region is excluded by those constraints.
We see that $\delta_{ct}$ is strongly constrained by the $B$ meson mixing and 
the magnitude of $\delta_{ct}$ should be less than $0.2 \, \pi$. 
Now, we fix $\rho_u^{tt}$ at unit, so that
our prediction is deviated from the SM prediction even if $\rho_u^{ct}$ is vanishing. 
Then, we find the allowed region for $\rho_u^{ct}$ around $0.1$. The right panel shows the
$m_H$ dependence on the flavor observables. If $m_H$ is heavier than 1 TeV, 
$|\rho_u^{ct}|$ can be large. On the other hand, the constraint from the $B_{d(s)}-\overline{B_{d(s)}}$ mixing
becomes severe if our charged scalar is light, as shown in the right panel of Fig. \ref{re2}. If we consider the charged scalar whose mass is less than 800 GeV, the enhancement of $\epsilon'/\epsilon$ becomes less than the one in the heavy scalar region in Fig. \ref{re2}.
Such a very light charged scalar case is actually discussed in Ref. \cite{Chen:2018ytc}, but we find that the region to enhance $\epsilon'/\epsilon$ is excluded by $\phi_{B_d}$ that is not analyzed in Ref. \cite{Chen:2018ytc}.

\begin{figure}[t]
  \begin{center}
                  \includegraphics[width=7.5cm]{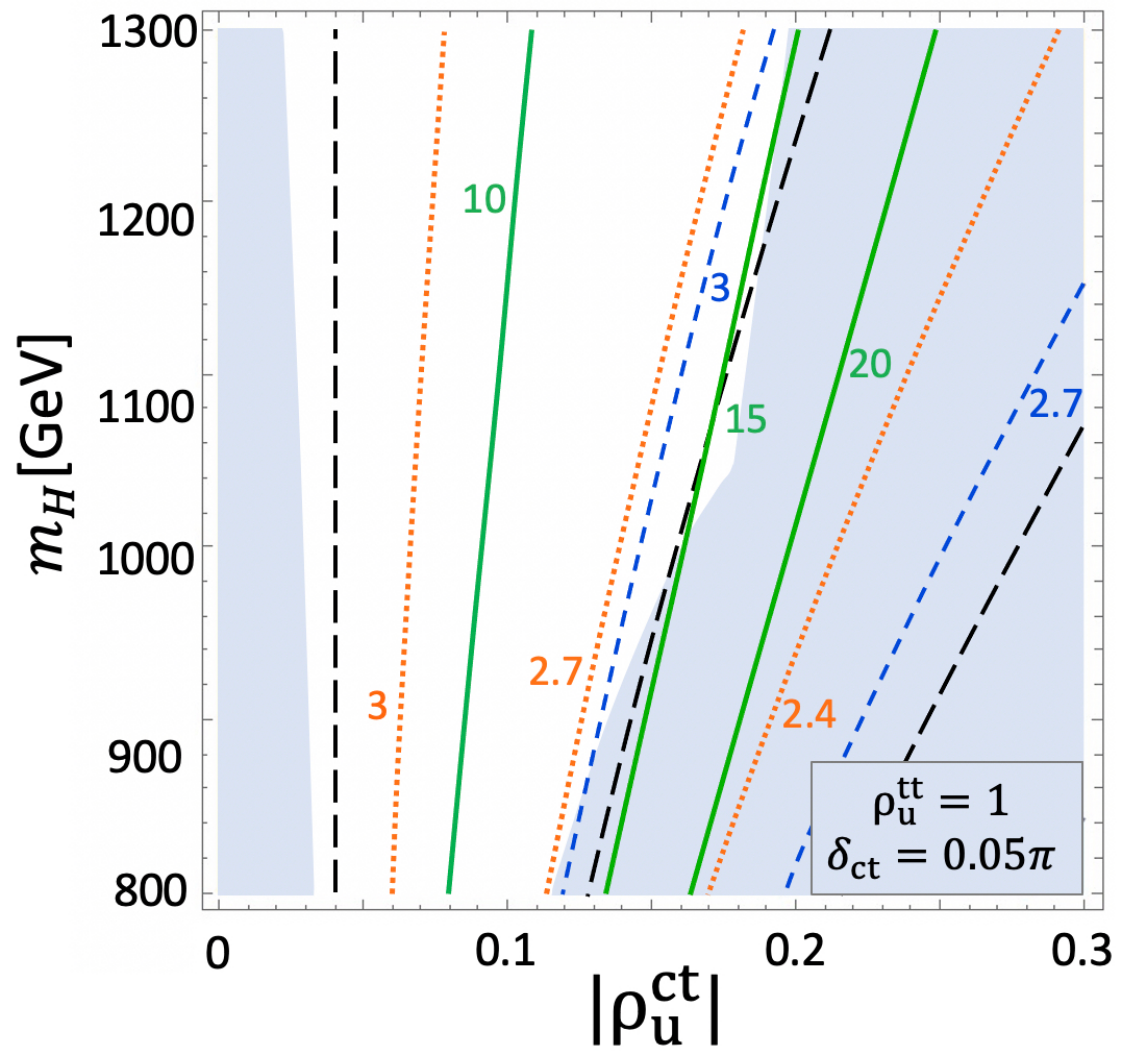}
        \caption{ Predictions for $BR(K^-\to \pi^-\nu\bar\nu)$, $BR(K^0\to \pi^0\nu\bar\nu)$ and $BR(B_S\to\mu\bar\mu)$. The green solid lines depict our predictions for $BR(K^+\to\pi^+ \nu\bar\nu)\times 10^{11}$. The orange solid lines express our predictions for $BR(K_L \to\pi^0 \nu\bar\nu)\times 10^{11}$.
The blue dashed lines are for $BR(B_s\to \mu\bar\mu)\times 10^{9}$. The black dashed lines are the same as in the right panel in Fig. \ref{re2}. }
    \label{re3}
  \end{center}
\end{figure}

Finally, we find that those observables strongly limit the enhancement of 
$(\epsilon'/\epsilon)$. In addition to the CKM matrix, there is another source of the CP violation, that is 
the phase of $\rho_u^{ct}$. The phase should be tuned to be consistent with especially $\phi_{B_d}$. Even if the phase is tuned, the enhancement of $\epsilon'/\epsilon$ is at most $3\times10^4$. 
The SM prediction has still ambiguity, so that we could not exclude our model but 
we conclude that our model cannot reach the $1 \sigma$ region of the experimental result if we use the SM prediction in Eq. (\ref{DQCD}).
   Note that our predictions of $BR(K\to\pi \nu\bar\nu)$ and $BR(B_s\to \mu\bar\mu)$ are deviated from the SM prediction in this case.
Fig. \ref{re3} summarizes our predictions on the same plane with the same dashed black lines as in the right panel of Fig. \ref{re2}.
The green solid lines correspond to our predictions for $BR(K^+\to\pi^+ \nu\bar\nu)\times 10^{11}$.
The orange dotted lines denote our predictions for $BR(K_L\to\pi^0 \nu\bar\nu)\times 10^{11}$.
The blue dashed lines are for $BR(B_s\to \mu\bar\mu)\times 10^{9}$.
Note that the SM predictions using our input parameters are $BR(K^+\to\pi^+ \nu\bar\nu)\times 10^{11}=9.0$, $BR(K_L\to\pi^0 \nu\bar\nu)\times 10^{11}=3.1$ and $BR(B_s\to \mu\bar\mu)\times 10^{9}=3.3$.
They are consistent with the results in Refs. \cite{Bobeth:2017ecx,Bobeth:2013uxa} within $1\sigma$ error. 
The shaded region is excluded by the constraints.
We find some correlations among the observables.
We see that $BR(K_L\to\pi^0 \nu\bar\nu)$ and $BR(B_s\to\mu\bar\mu)$ are suppressed if $\left( \epsilon'/\epsilon \right )_H$ is large, while $BR(K^-\to \pi^-\nu\bar\nu)$ is enhanced. The deviation reaches about \si{50} \% of the SM prediction
when the enhancement of $\epsilon'/\epsilon$ is maximum. 

\begin{figure}[t]
  \begin{center}
     \includegraphics[width=8cm]{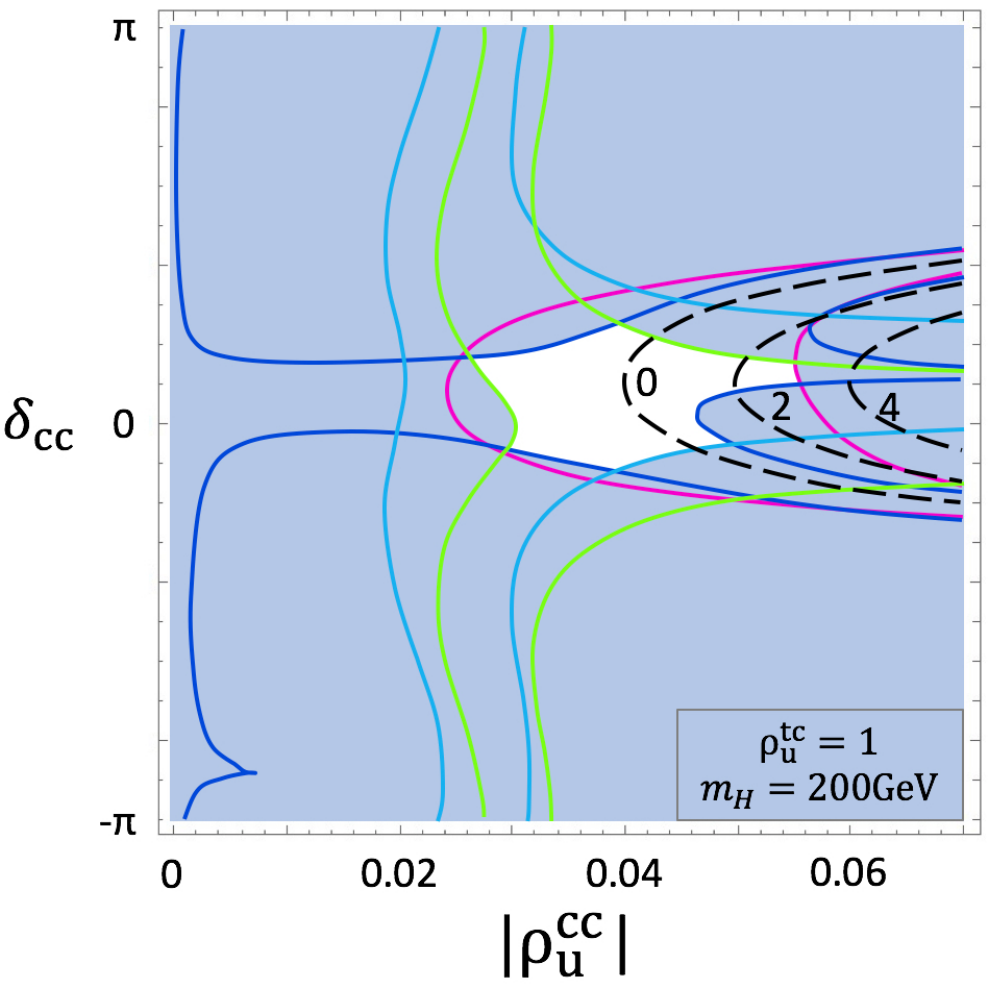}
            \caption{ The prediction for $\epsilon'/\epsilon_{\rm{H}}\times10^4$ is expressed in the case (II) with the $\delta_{cc}$ and $\left | \rho_u^{cc} \right |$ plane setting $ \rho_u^{tc}=1 $ and $m_H=200$ GeV. The white region is allowed by constraints. The color code for the prediction and constraints is the same as in the left of Fig. \ref{re2}  }
    \label{re1}
  \end{center}
\end{figure}

Next, we consider the case (II). In this case, the loop correction involving the charged scalar and charm quark
is dominant and it is linear to $\rho_u^{tc}\times\rho_u^{cc}$.
The relevant parameters are $m_H$, $|\rho_u^{tc}|$, $|\rho_u^{cc}|$ and a relative phase between $\rho_u^{tc}$ and $\rho_u^{cc}$.
Now we take $\rho_u^{tc}$ to be real and $\rho_u^{cc}$ to be complex without a loss of generality.
This loop is enhanced by $\log (m_c/m_H)$ in Eq. (26), but the loop correction is small unless the charged scalar is light.
If $m_H<m_t$, $\rho_u^{tc}$ generates a flavor changing top decay as $t\to Hc$ and $t\to Ac$, so that we fix $m_H$ at $m_{H}=200$ GeV.
 We plot our prediction of $(\epsilon'/\epsilon)_{{\rm H}} \times 10^4$ with dashed black lines in Fig. \ref{re1}.
 We fixed $\rho_u^{tc}=1$ to draw the figure.
 The gray shaded region is excluded by the various constraints as discussed in the previous section.
 The vertical and horizontal axes correspond to the phase and the absolute value of $\rho_u^{cc}$. We define the phase as $\rho_u^{cc}=|\rho_u^{cc}| e^{i\delta_{cc}}$.
 In this case, $\delta_{cc}$ should be smaller than $0.2 \, \pi$ and the enhancement of $\epsilon'/\epsilon$ can be $3\times10^4$.
On the other hand, the dominant contribution to $\epsilon'/\epsilon$ is from the photon penguin diagram in Fig. \ref {diaBP} in the case (II), while the contribution to $BR(K\to\pi \nu\bar\nu)$ and $BR(B_s\to \mu\bar\mu)$ comes from the Z-penguin diagram that is subdominant in  $(\epsilon'/\epsilon)_{{\rm H}}$.
Thus the deviation from the SM predictions in $BR(K\to\pi \nu\bar\nu)$ and $BR(B_s\to \mu\bar\mu)$ is less than $1\%$ in the case (II).


\subsection{The collider constraint and prediction}
\label{sec4}
Before closing this paper, we propose processes to test our scenario at the LHC in both cases.  

In the case (I), $\rho_u^{tt}$ is large, so that we expect some signal associated with top quarks.
There are several channel to test our model; for instance, $4t$ and $2t$ in the final states.
We calculate our signal using Madgraph5 \cite{Alwall:2011uj} with $\sqrt s=13$ TeV.
In this case, $\rho_u^{tt}$ is much larger than $\rho_u^{ct}$ so that we assume $BR(H\to t\bar t)=BR(A\to t\bar t)\sim1$ for simplicity.
\begin{figure}[t]
  \begin{center}
      \includegraphics[width=8cm]{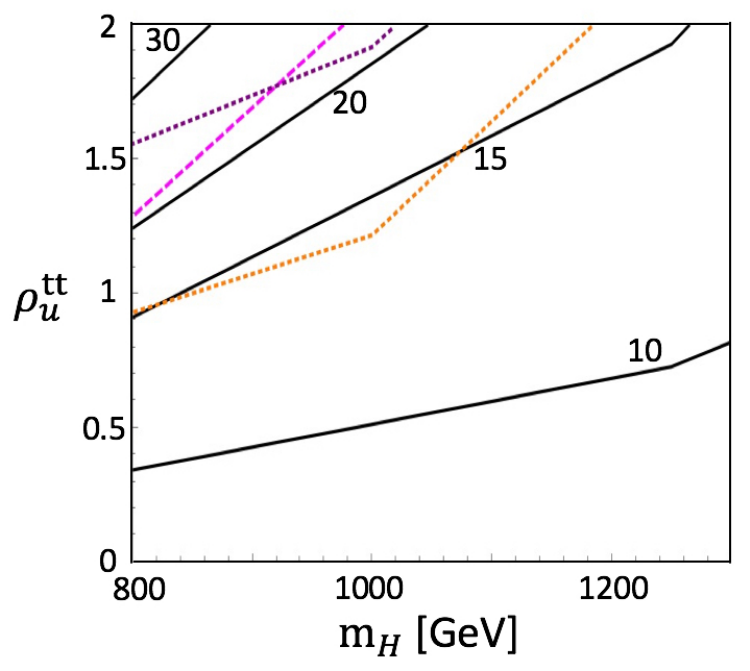}
        \caption{ Our prediction for 4 top in the final state at the LHC in the case (I).
        The vertical and horizontal axes denote the heavy Higgs mass and $\rho_u^{tt}$.
        The solid black lines are our prediction of the cross section including the SM contribution for the signal with $\sqrt s=13$ TeV. 
The numbers attached to the black lines indicate the cross sections of $pp \to t t \overline{t} \overline{t}$ in $fb$.
A magenta dashed line expresses the upper limit from the $t\bar{t}$ resonance search reported by the ATLAS.
A purple (orange) dotted line is obtained from the constraint with the assumption, $\Gamma_{Z^\prime}=0.1(0.01)m_{Z^\prime}$ by the CMS.
See the main text for the further description.          }
    \label{FC2}
  \end{center}
\end{figure}

We plot our prediction for the $4t$ signal in Fig. \ref{FC2}. 
The solid black lines are our prediction of the cross section including the SM contribution for the signal with 4 top in the final state. 
The numbers attached to the black lines indicate the cross sections of $pp \to t t \overline{t} \overline{t}$ in $fb$.
The SM prediction is 9.2 $fb$.
Currently a mild excess is reported by the ATLAS collaboration as 
 $\mu_C=3.1^{+1.3}_{-1.2}$ \cite{Aaboud:2018jsj}, where $\mu_C$ is defined by the ratio of cross sections as $\mu_C = \sigma_{exp}/\sigma_{SM}$.
The CMS collaboration also announces the result as 
  $\mu_C=1.8^{+1.5}_{-1.3}$ \cite{Sirunyan:2017roi}.

We draw the bound from the $t\bar{t}$ resonance search. We apply the constraint for a heavy vector boson, namely $Z'$, assuming $m_A=m_H$.
In Fig. \ref{FC2} a magenta dashed line expresses the upper limit from the $t\bar{t}$ resonance search reported by the ATLAS at 95$\%$ CL. \cite{Aaboud:2018mjh}. 
Note that the bound from a charged Higgs search: $H^-\to \bar{t}b$ search\cite{ATLAS:2016qiq} is weaker than the one in Ref. \cite{Aaboud:2018mjh}.
There is also a result on the $t\bar{t}$ heavy resonance search reported by the CMS collaboration \cite{Sirunyan:2018ryr}.
A purple dotted line is obtained from the constraint with the assumption, $\Gamma_{Z^\prime}=0.1m_{Z^\prime}$ at 95$\%$ CL.
We also plot the orange dotted line that is from the constraint with the assumption, $\Gamma_{Z^\prime}=0.01m_{Z^\prime}$ at 95$\%$ CL.
Note that when particles in decay of heavy scalars $H$ and $A$ is light enough in comparison with them, the decay width scalar is given as $\Gamma_{A(H)}\sim0.06|\rho_u^{tt}|^2m_{A(H)}$.
Therefore, applying the bound of the orange line seems to be too aggressive.
In Fig. \ref{FC2}, we see that $\mu_C$ could be about $2$, and both 4 top search and $t\bar{t}$ resonance search with Run 2 full data may test our scenario e.g. $m_H=1000$GeV and $\rho^{tt}_u\simeq 1$.

The collider physics of the case (II) with sizable $\rho_u^{tc}$ and $\rho_u^{cc}$ needs a discussion.
 In this scenario, $\rho_u^{tc}$ is much larger than $\rho_u^{cc}$. 
 The analysis in Ref. \cite{Iguro:2018qzf} can be applied to our model.
 The flavor inclusive dijet search result to look for the charged scalar that decays dominantly to $bc$ is available for more than $450$ GeV \cite{Aaboud:2018fzt}.
An additional b tagged jet can be helpful to look into the lighter mass resonance in the future.
 If CP even and odd scalars are not degenerated, a large $\rho_u^{tc}$ would generate the same-sign top signal \cite{Hou:2018zmg}.

\section{Summary}
\label{summary and comment}
We have studied the phenomenology in a general 2HDM with tree-level FCNCs. 
In the bottom-up approach, there are many free parameters in the Yukawa couplings between the heavy scalars and the SM fermions. The parameters may be fixed by the underlying theory, but there are actually many candidates for the extended SM behind this kind of model. Then, it would be important to survey the experimental constraints on the each coupling in this model as well, not specifying the underlying theory.

Most of the parameters are strongly constrained by flavor physics. The only Yukawa couplings that can be sizable would be the ones involving heavy quarks: charm and top quarks. The bounds are not relatively tight, so that
we can expect that the signals caused by the couplings and the heavy scalar are observed in the LHC and the flavor experiments. In this paper, we have focused on the physics caused by the couplings between the heavy scalar and heavy quarks. 

One of the flavor observables attracting attention recently is $\epsilon'/\epsilon$ related to 
the direct CP violation of $K \to \pi \pi$.
Both of the direct and indirect CP violations in the $K \to \pi \pi$ process are experimentally measured 
with high accuracy. The SM prediction, on the other hand, suffers from the large theoretical uncertainty.
Recently, the calculation of the hadronic process has been done by the RBC-UKQCD lattice collaboration
and we find the discrepancy between the SM prediction and the experimental result.
Motivated by the excess, some new physics possibilities have been discussed, and
one good candidate is the extended SM with extra Higgs doublets that have large FCNCs involving heavy quarks \cite{Chen:2018ytc,Chen:2018vog,Marzo:2019ldg}.
We find that the couplings can enhance the direct CP violation, but
the constraint from the CP violation in the $B$ decay is so strong that the enhancement is at most 
$\epsilon'/\epsilon_{\rm{H}} \approx 3 \times10^{-4}$.
This can explain the discrepancy of the direct CP violation at the $2 \sigma$ level.
We have considered the two cases: the top loop is dominant in the case (I) and the charm loop is dominant in the case (II). The favored mass region for the heavy scalar is different because of the quark mass dependence on loop functions.
$m_H$ should be about 1 TeV in the case (I), while $m_H$ should be about 200 GeV in the case (II).
In Ref. \cite{Chen:2018ytc,Chen:2018vog}, the authors concluded that a general 2HDM can enhance the $\epsilon^\prime/\epsilon$ by considering the light charged scalar and the cromomagnetic operator with the large mass difference between $m_{H^-}$ and $m_H=m_A$. 
In our result, 
the light scalar region is excluded by the constraint from the CP violation of the $B$ decay. We also confirm that
the contribution of the cromomagnetic operator is very small in the allowed region.
We also note that the parameter set to enhance $\epsilon^\prime/\epsilon$, that is introduced in Refs. \cite{Chen:2018ytc,Chen:2018vog}, is excluded in our analysis, taking into account the electroweak precision observables as well.

Finally, we have discussed how to test our scenario at the LHC. Our setup predicts the 4 top signal at the LHC, so that we have estimated the cross section of $pp \to t \overline{t}t \overline{t}$.
The enhancement of the direct CP requires $\rho_u^{tt}$ to be ${\cal O}(1)$ so
we may be able to test our model if $m_H$ is about 1 TeV, according to Fig. \ref{FC2}.

\section*{Acknowledegments}
We would like to acknowledge Kei Yamamoto for a collaboration at the early stage.
We also thank Girish Kumar for pointing out the error in Eq.\,(35).
The work of S. I. is supported by Kobayashi-Maskawa Institute for the Origin of Particles and the Universe, Toyoaki scholarship foundation and the Japan Society for the Promotion of Science (JSPS) Research Fellowships for Young Scientists, No. 19J10980.
The work of Y. O. is supported by Grant-in-Aid for Scientific research from the Ministry of Education, Science, Sports, and Culture (MEXT), Japan, No. 19H04614, No. 19H05101 and No. 19K03867.



\begin{thebibliography}{99}
{\small

\bibitem{GUT}
H.~Georgi and S.~L.~Glashow, 
Phys.\ Rev. \ Lett. {\bf 32}, 438 (1974);
  H.~Georgi,
  AIP Conf.\ Proc.\  {\bf 23}, 575 (1975);
  H.~Fritzsch and P.~Minkowski,
  Annals Phys.\  {\bf 93}, 193 (1975).

\bibitem{Babu:1989rb} 
  K.~S.~Babu and R.~N.~Mohapatra,
  Phys.\ Rev.\ D {\bf 41}, 1286 (1990).

\bibitem{Huitu:1994zm} 
  K.~Huitu and J.~Maalampi,
  Phys.\ Lett.\ B {\bf 344}, 217 (1995)
  [hep-ph/9410342].
 
\bibitem{Frank:2014kma} 
  M.~Frank, D.~K.~Ghosh, K.~Huitu, S.~K.~Rai, I.~Saha and H.~Waltari,
  Phys.\ Rev.\ D {\bf 90}, no. 11, 115021 (2014)
  [arXiv:1408.2423 [hep-ph]].
 
  \bibitem{Babu:2014vba} 
  K.~S.~Babu and A.~Patra,
  Phys.\ Rev.\ D {\bf 93}, no. 5, 055030 (2016)
  [arXiv:1412.8714 [hep-ph]].
  
\bibitem{Frank:2011jia} 
  M.~Frank and B.~Korutlu,
  Phys.\ Rev.\ D {\bf 83}, 073007 (2011)
  [arXiv:1101.3601 [hep-ph]].

\bibitem{Dev:2016dja} 
  P.~S.~B.~Dev, R.~N.~Mohapatra and Y.~Zhang,
  JHEP {\bf 1605}, 174 (2016)
  [arXiv:1602.05947 [hep-ph]].
  
\bibitem{Iguro:2018oou} 
  S.~Iguro, Y.~Muramatsu, Y.~Omura and Y.~Shigekami,
  JHEP {\bf 1811}, 046 (2018)
  [arXiv:1804.07478 [hep-ph]].
  
\bibitem{Lees2012xj} 
  J.~P.~Lees {\it et al.} [BaBar Collaboration],
  Phys.\ Rev.\ Lett.\  {\bf 109}, 101802 (2012)
  [arXiv:1205.5442 [hep-ex]].
  

\bibitem{Lees2013uzd} 
  J.~P.~Lees {\it et al.} [BaBar Collaboration],
  Phys.\ Rev.\ D {\bf 88}, no. 7, 072012 (2013)
  [arXiv:1303.0571 [hep-ex]].
    
\bibitem{Aaij2015yra} 
  R.~Aaij {\it et al.} [LHCb Collaboration],
  Phys.\ Rev.\ Lett.\  {\bf 115}, no. 11, 111803 (2015)
  [arXiv:1506.08614 [hep-ex]].

\bibitem{Huschle2015rga} 
  M.~Huschle {\it et al.} [Belle Collaboration],
  Phys.\ Rev.\ D {\bf 92}, no. 7, 072014 (2015)
  [arXiv:1507.03233 [hep-ex]].

\bibitem{Sato2016svk} 
  Y.~Sato {\it et al.} [Belle Collaboration],
  Phys.\ Rev.\ D {\bf 94}, no. 7, 072007 (2016)
  [arXiv:1607.07923 [hep-ex]].
  

\bibitem{Hirose2016wfn} 
  S.~Hirose {\it et al.} [Belle Collaboration],
  Phys.\ Rev.\ Lett.\  {\bf 118}, no. 21, 211801 (2017)
  [arXiv:1612.00529 [hep-ex]].
  
\bibitem{Aaij:2013qta} 
  R.~Aaij {\it et al.} [LHCb Collaboration],
  Phys.\ Rev.\ Lett.\  {\bf 111}, 191801 (2013)
  [arXiv:1308.1707 [hep-ex]].

\bibitem{Aaij:2015oid} 
  R.~Aaij {\it et al.} [LHCb Collaboration],
  JHEP {\bf 1602}, 104 (2016)
  [arXiv:1512.04442 [hep-ex]].
 
  \bibitem{LHCbnew}
  R.~Aaij {\it et al.} [LHCb Collaboration],
  JHEP {\bf 1708}, 055 (2017)
  [arXiv:1705.05802 [hep-ex]].

\bibitem{Aaij:2014ora} 
  R.~Aaij {\it et al.} [LHCb Collaboration],
  Phys.\ Rev.\ Lett.\  {\bf 113}, 151601 (2014)
  [arXiv:1406.6482 [hep-ex]].
  
  \bibitem{Aaij:2019wad} 
  R.~Aaij {\it et al.} [LHCb Collaboration],
  Phys.\ Rev.\ Lett.\  {\bf 122}, no. 19, 191801 (2019)
  [arXiv:1903.09252 [hep-ex]].
  \bibitem{Abdesselam:2019wac} 
  A.~Abdesselam {\it et al.} [Belle Collaboration],
  arXiv:1904.02440 [hep-ex].
\bibitem{Abdesselam:2019dgh} 
  A.~Abdesselam {\it et al.} [Belle Collaboration],
  arXiv:1904.08794 [hep-ex].

  
\bibitem{Crivellin2012ye} 
  A.~Crivellin, C.~Greub and A.~Kokulu,
  Phys.\ Rev.\ D {\bf 86}, 054014 (2012)
  [arXiv:1206.2634 [hep-ph]].

  \bibitem{Celis:2012dk} 
  A.~Celis, M.~Jung, X.~Q.~Li and A.~Pich,
  JHEP {\bf 1301}, 054 (2013)
  [arXiv:1210.8443 [hep-ph]].
  
\bibitem{Tanaka2012nw} 
  M.~Tanaka and R.~Watanabe,
  Phys.\ Rev.\ D {\bf 87}, no. 3, 034028 (2013)
  [arXiv:1212.1878 [hep-ph]].

  
\bibitem{Crivellin:2013wna} 
  A.~Crivellin, A.~Kokulu and C.~Greub,
  Phys.\ Rev.\ D {\bf 87}, no. 9, 094031 (2013)
  [arXiv:1303.5877 [hep-ph]].
  
  
  \bibitem{Crivellin:2015hha} 
  A.~Crivellin, J.~Heeck and P.~Stoffer,
  Phys.\ Rev.\ Lett.\  {\bf 116}, no. 8, 081801 (2016)
  [arXiv:1507.07567 [hep-ph]].
  
\bibitem{Cline:2015lqp}
  J.~M.~Cline,
  Phys.\ Rev.\ D {\bf 93}, no. 7, 075017 (2016)
  [arXiv:1512.02210 [hep-ph]].


\bibitem{Ko:2017lzd} 
  P.~Ko, Y.~Omura, Y.~Shigekami and C.~Yu,
  Phys.\ Rev.\ D {\bf 95}, no. 11, 115040 (2017)
  [arXiv:1702.08666 [hep-ph]].
  
   
\bibitem{Ko:2012sv} 
  P.~Ko, Y.~Omura and C.~Yu,
  JHEP {\bf 1303}, 151 (2013)
  [arXiv:1212.4607 [hep-ph]].
  
\bibitem{Iguro:2017ysu} 
  S.~Iguro and K.~Tobe,
  Nucl.\ Phys.\ B {\bf 925}, 560 (2017)
  [arXiv:1708.06176 [hep-ph]].

  
\bibitem{Bian:2017rpg}
  L.~Bian, S.~M.~Choi, Y.~J.~Kang and H.~M.~Lee,
  Phys.\ Rev.\ D {\bf 96} (2017) no.7,  075038
  [arXiv:1707.04811 [hep-ph]].
  
\bibitem{Bian:2017xzg} 
  L.~Bian, H.~M.~Lee and C.~B.~Park,
  Eur.\ Phys.\ J.\ C {\bf 78}, no. 4, 306 (2018)
  [arXiv:1711.08930 [hep-ph]].

\bibitem{Martinez:2018ynq} 
  R.~Martinez, C.~F.~Sierra and G.~Valencia,
  Phys.\ Rev.\ D {\bf 98}, no. 11, 115012 (2018)
  [arXiv:1805.04098 [hep-ph]].

\bibitem{Iguro:2018fni} 
  S.~Iguro, Y.~Omura and M.~Takeuchi,
  Phys.\ Rev.\ D {\bf 99}, no. 7, 075013 (2019)
  [arXiv:1810.05843 [hep-ph]].




\bibitem{Hu:2016gpe} 
  Q.~Y.~Hu, X.~Q.~Li and Y.~D.~Yang,
  Eur.\ Phys.\ J.\ C {\bf 77}, no. 3, 190 (2017)
  [arXiv:1612.08867 [hep-ph]].
  
\bibitem{Arnan:2017lxi} 
  P.~Arnan, D.~Be\v{c}irevi\'c, F.~Mescia and O.~Sumensari,
  Eur.\ Phys.\ J.\ C {\bf 77}, no. 11, 796 (2017)
  [arXiv:1703.03426 [hep-ph]].
\bibitem{Li:2018rax} 
  S.~P.~Li, X.~Q.~Li, Y.~D.~Yang and X.~Zhang,
  JHEP {\bf 1809}, 149 (2018)
  [arXiv:1807.08530 [hep-ph]].
  
\bibitem{Iguro:2018qzf} 
  S.~Iguro and Y.~Omura,
  JHEP {\bf 1805}, 173 (2018)
  [arXiv:1802.01732 [hep-ph]].
  
  


  
\bibitem{Omura:2015nja} 
  Y.~Omura, E.~Senaha and K.~Tobe,
  JHEP {\bf 1505}, 028 (2015)
  [arXiv:1502.07824 [hep-ph]].
  
\bibitem{Omura:2015xcg} 
  Y.~Omura, E.~Senaha and K.~Tobe,
  Phys.\ Rev.\ D {\bf 94}, no. 5, 055019 (2016)
  [arXiv:1511.08880 [hep-ph]].

\bibitem{IOT2}
  S.~Iguro, Y.~Omura and M.~Takeuchi,
  arXiv:1907.09845 [hep-ph].

\bibitem{Cheng:1987rs} 
  T.~P.~Cheng and M.~Sher,
  Phys.\ Rev.\ D {\bf 35}, 3484 (1987).
  
\bibitem{Babu:2018uik} 
  K.~S.~Babu and S.~Jana,
  JHEP {\bf 1902}, 193 (2019)
  [arXiv:1812.11943 [hep-ph]].

\bibitem{Batley:2002gn} 
  J.~R.~Batley {\it et al.} [NA48 Collaboration],
  Phys.\ Lett.\ B {\bf 544}, 97 (2002)
  [hep-ex/0208009].
\bibitem{AlaviHarati:2002ye} 
  A.~Alavi-Harati {\it et al.} [KTeV Collaboration],
  Phys.\ Rev.\ D {\bf 67}, 012005 (2003)
  [hep-ex/0208007].
\bibitem{Abouzaid:2010ny} 
  E.~Abouzaid {\it et al.} [KTeV Collaboration],
  Phys.\ Rev.\ D {\bf 83}, 092001 (2011)
  [arXiv:1011.0127 [hep-ex]].

\bibitem{Buras:2015yba} 
  A.~J. Buras, M.~Gorbahn, S.~Jager and M.~Jamin,
  JHEP {\bf 1511}, 202 (2015)
  [arXiv:1507.06345 [hep-ph]].

\bibitem{Kitahara:2016nld} 
  T.~Kitahara, U.~Nierste and P.~Tremper,
  JHEP {\bf 1612}, 078 (2016)
  [arXiv:1607.06727 [hep-ph]].

\bibitem{Blum:2015ywa} 
  T.~Blum {\it et al.},
  Phys.\ Rev.\ D {\bf 91}, no. 7, 074502 (2015)
  [arXiv:1502.00263 [hep-lat]].

\bibitem{Bai:2015nea} 
  Z.~Bai {\it et al.} [RBC and UKQCD Collaborations],
  Phys.\ Rev.\ Lett.\  {\bf 115}, no. 21, 212001 (2015)
  [arXiv:1505.07863 [hep-lat]].
   
   \bibitem{Buras:2015xba} 
  A.~J.~Buras and J.~M.~Gerard,
  JHEP {\bf 1512}, 008 (2015)
  [arXiv:1507.06326 [hep-ph]].
  
\bibitem{Buras:2016fys} 
  A.~J.~Buras and J.~M.~Gerard,
  Eur.\ Phys.\ J.\ C {\bf 77}, no. 1, 10 (2017)
  [arXiv:1603.05686 [hep-ph]].

\bibitem{Gisbert:2017vvj} 
  H.~Gisbert and A.~Pich,
  Rept.\ Prog.\ Phys.\  {\bf 81}, no. 7, 076201 (2018)
  [arXiv:1712.06147 [hep-ph]].
  
\bibitem{Aebischer:2018csl} 
  J.~Aebischer, C.~Bobeth, A.~J.~Buras and D.~M.~Straub,
  Eur.\ Phys.\ J.\ C {\bf 79}, no. 3, 219 (2019)
  [arXiv:1808.00466 [hep-ph]].
  
\bibitem{Buras:2015kwd} 
  A.~J.~Buras and F.~De Fazio,
  JHEP {\bf 1603}, 010 (2016)
  [arXiv:1512.02869 [hep-ph]].
  
\bibitem{Tanimoto:2016yfy} 
  M.~Tanimoto and K.~Yamamoto,
  PTEP {\bf 2016}, no. 12, 123B02 (2016)
  [arXiv:1603.07960 [hep-ph]].
  
\bibitem{Buras:2016dxz} 
  A.~J.~Buras and F.~De Fazio,
  JHEP {\bf 1608}, 115 (2016)
  [arXiv:1604.02344 [hep-ph]].
 
\bibitem{Endo:2016aws} 
  M.~Endo, S.~Mishima, D.~Ueda and K.~Yamamoto,
  Phys.\ Lett.\ B {\bf 762}, 493 (2016)
  [arXiv:1608.01444 [hep-ph]].
  
\bibitem{Bobeth:2016llm} 
  C.~Bobeth, A.~J.~Buras, A.~Celis and M.~Jung,
  JHEP {\bf 1704}, 079 (2017)
  [arXiv:1609.04783 [hep-ph]].
  
\bibitem{Endo:2016tnu} 
  M.~Endo, T.~Kitahara, S.~Mishima and K.~Yamamoto,
  Phys.\ Lett.\ B {\bf 771}, 37 (2017)
  [arXiv:1612.08839 [hep-ph]].
 
\bibitem{Bobeth:2017xry} 
  C.~Bobeth, A.~J.~Buras, A.~Celis and M.~Jung,
  JHEP {\bf 1707}, 124 (2017)
  [arXiv:1703.04753 [hep-ph]].
  
\bibitem{Crivellin:2017gks} 
  A.~Crivellin, G.~D'Ambrosio, T.~Kitahara and U.~Nierste,
  Phys.\ Rev.\ D {\bf 96}, no. 1, 015023 (2017)
  [arXiv:1703.05786 [hep-ph]].
  
\bibitem{Haba:2018byj} 
  N.~Haba, H.~Umeeda and T.~Yamada,
  JHEP {\bf 1805}, 052 (2018)
  [arXiv:1802.09903 [hep-ph]].
  
\bibitem{Haba:2018rzf} 
  N.~Haba, H.~Umeeda and T.~Yamada,
  JHEP {\bf 1810}, 006 (2018)
  [arXiv:1806.03424 [hep-ph]].
  
  
  \bibitem{Altunkaynak:2015twa} 
  B.~Altunkaynak, W.~S.~Hou, C.~Kao, M.~Kohda and B.~McCoy,
  Phys.\ Lett.\ B {\bf 751}, 135 (2015)
  [arXiv:1506.00651 [hep-ph]].

\bibitem{Chen:2018ytc} 
  C.~H.~Chen and T.~Nomura,
  JHEP {\bf 1808}, 145 (2018)
  [arXiv:1804.06017 [hep-ph]].


\bibitem{Georgi:1978ri} 
  H.~Georgi and D.~V.~Nanopoulos,
  Phys.\ Lett.\  {\bf 82B}, 95 (1979).
  
\bibitem{Donoghue:1978cj} 
  J.~F.~Donoghue and L.~F.~Li,
  Phys.\ Rev.\ D {\bf 19}, 945 (1979).

\bibitem{Peskin:1990zt} 
  M.~E.~Peskin and T.~Takeuchi,
  Phys.\ Rev.\ Lett.\  {\bf 65}, 964 (1990).

\bibitem{Buras:1993dy} 
  A.~J.~Buras, M.~Jamin and M.~E.~Lautenbacher,
  Nucl.\ Phys.\ B {\bf 408}, 209 (1993)
  [hep-ph/9303284].

\bibitem{Buchalla:1995vs}
  G.~Buchalla, A.~J.~Buras and M.~E.~Lautenbacher,
  Rev.\ Mod.\ Phys.\  {\bf 68} (1996) 1125
  [hep-ph/9512380].

\bibitem{Buras:2015jaq} 
  A.~J.~Buras,
  JHEP {\bf 1604}, 071 (2016)
  [arXiv:1601.00005 [hep-ph]].
  
\bibitem{Cirigliano:2003nn} 
  V.~Cirigliano, A.~Pich, G.~Ecker and H.~Neufeld,
  Phys.\ Rev.\ Lett.\  {\bf 91}, 162001 (2003)
  [hep-ph/0307030].
  
\bibitem{Chen:2018vog} 
  C.~H.~Chen and T.~Nomura,
  Phys.\ Lett.\ B {\bf 787}, 182 (2018)
  [arXiv:1805.07522 [hep-ph]].
  


\bibitem{UTfit}
  M.~Bona {\it et al.} [UTfit Collaboration],
  JHEP {\bf 0803}, 049 (2008)
  [arXiv:0707.0636 [hep-ph]].
see http://www.utfit.org/UTfit/ResultsSummer2018NP for update.
  
\bibitem{Tanabashi:2018oca} 
  M.~Tanabashi {\it et al.} [Particle Data Group],
  Phys.\ Rev.\ D {\bf 98}, no. 3, 030001 (2018).
  
\bibitem{Isidori:2003ts} 
  G.~Isidori and R.~Unterdorfer,
  JHEP {\bf 0401}, 009 (2004)
  [hep-ph/0311084].
  
  \bibitem{Buras:2015yca} 
  A.~J.~Buras, D.~Buttazzo and R.~Knegjens,
  JHEP {\bf 1511}, 166 (2015)
  [arXiv:1507.08672 [hep-ph]].
  
\bibitem{Lurkin:2018gdo} 
  N.~Lurkin [NA62 Collaboration],
  EPJ Web Conf.\  {\bf 199}, 01007 (2019)
  [arXiv:1809.05384 [hep-ex]].
  
\bibitem{Buchalla:1998ba} 
  G.~Buchalla and A.~J.~Buras,
  Nucl.\ Phys.\ B {\bf 548}, 309 (1999)
  [hep-ph/9901288].
  
  \bibitem{Buras:2015qea} 
  A.~J.~Buras, D.~Buttazzo, J.~Girrbach-Noe and R.~Knegjens,
  JHEP {\bf 1511}, 033 (2015)
  [arXiv:1503.02693 [hep-ph]].


  

 
\bibitem{Bobeth:2013uxa} 
  C.~Bobeth, M.~Gorbahn, T.~Hermann, M.~Misiak, E.~Stamou and M.~Steinhauser,
  Phys.\ Rev.\ Lett.\  {\bf 112}, 101801 (2014)
  [arXiv:1311.0903 [hep-ph]].

\bibitem{Bobeth:2017ecx} 
  C.~Bobeth and A.~J.~Buras,
  JHEP {\bf 1802}, 101 (2018)
  [arXiv:1712.01295 [hep-ph]].
  
 
  

\bibitem{Alwall:2011uj} 
  J.~Alwall, M.~Herquet, F.~Maltoni, O.~Mattelaer and T.~Stelzer,
  JHEP {\bf 1106}, 128 (2011)
  [arXiv:1106.0522 [hep-ph]].

\bibitem{Aaboud:2018jsj} 
  M.~Aaboud {\it et al.} [ATLAS Collaboration],
  Phys.\ Rev.\ D {\bf 99}, no. 5, 052009 (2019)
  [arXiv:1811.02305 [hep-ex]].

\bibitem{Sirunyan:2017roi} 
  A.~M.~Sirunyan {\it et al.} [CMS Collaboration],
  Eur.\ Phys.\ J.\ C {\bf 78}, no. 2, 140 (2018)
  [arXiv:1710.10614 [hep-ex]].
  
\bibitem{Aaboud:2018mjh}
  M.~Aaboud {\it et al.} [ATLAS Collaboration],
  Eur.\ Phys.\ J.\ C {\bf 78} (2018) no.7,  565
  [arXiv:1804.10823 [hep-ex]].
 
 
\bibitem{ATLAS:2016qiq} 
  The ATLAS collaboration [ATLAS Collaboration],
  ATLAS-CONF-2016-089.

 
\bibitem{Sirunyan:2018ryr} 
  A.~M.~Sirunyan {\it et al.} [CMS Collaboration],
  JHEP {\bf 1904}, 031 (2019)
  [arXiv:1810.05905 [hep-ex]].
  
  \bibitem{Aaboud:2018fzt} 
  M.~Aaboud {\it et al.} [ATLAS Collaboration],
  Phys.\ Rev.\ Lett.\  {\bf 121}, no. 8, 081801 (2018)
  [arXiv:1804.03496 [hep-ex]].

  
  \bibitem{Hou:2018zmg} 
  W.~S.~Hou, M.~Kohda and T.~Modak,
  Phys.\ Lett.\ B {\bf 786}, 212 (2018)
  [arXiv:1808.00333 [hep-ph]].
  
\bibitem{Marzo:2019ldg} 
  C.~Marzo, L.~Marzola and M.~Raidal,
  arXiv:1901.08290 [hep-ph].
  

 
  
}
\end{thebibliography}
\end{document}